\documentclass[aps,pra,showpacs,twoside,twocolumn,10pt]{revtex4-1}
\usepackage[colorlinks=true, citecolor=red, urlcolor=blue ]{hyperref}
\usepackage{epsfig,newlfont,amssymb,amsfonts,amsmath,bm,subfigure,palatino,mathtools,amsthm,braket,soul,enumitem,color,graphics,graphicx,times,physics,bbold}
\usepackage[normalem]{ulem}
\usepackage{xcolor}
\usepackage{physics}
\usepackage{dsfont}
\usepackage{mathrsfs}
\usepackage{verbatim}
\usepackage{amsthm,amssymb}

\begin{document}
\title{
Resource theory of nonabsolute separability 
}

\author{Ayan Patra, Arghya Maity, Aditi Sen(De)}
\affiliation{Harish-Chandra Research Institute, A CI of Homi Bhabha National
Institute,  Chhatnag Road, Jhunsi, Allahabad - 211019, India}

\begin{abstract}

We develop a resource theory for non-absolutely separable states (non-AS) in which absolutely separable states (AS) that cannot be entangled by any global unitaries are recognized as free states and any convex mixture of global unitary operations can be performed without incurring any costs. We employ two approaches to quantify non-absolute separability (NAS) --  one  based on distance measures and the other one through the use of a witness operator. We prove that both of the NAS measures obey all the conditions that should be followed by a ``good'' NAS measure. We demonstrate that NAS content is equal and maximal in all pure states  for a fixed dimension. We then establish a connection between the distance-based NAS measure and the entanglement quantifier. We illustrate our results with a class of non-AS states, namely Werner states.

\end{abstract}

\maketitle

\section{Introduction}
\label{sec:intro}

The physical processes that are permitted under a theory are determined by some physical rules. 
  For instance,  some no-go theorems in quantum mechanics, such as no-cloning \cite{Wootters_1982, Dieks_PLA_1982}, no-deleting \cite{Arun00}, no-broadcasting \cite{Barnum96} theorems were discovered since unitary operations do not permit these processes to take place.  Under quantum mechanics, if one confines the class of operations, referred to as free operations, one can identify a set of states, called a resource  that cannot be prepared with free operations, \cite{Horodecki_IJMPB_2013,Chitambar_RMP_2019}. According to this idea of resources, some protocols that surpass their classical counterparts can only be designed utilizing resourceful quantum states. It is fascinating to learn how well the  resource theory can perform particular jobs when presented with a particular collection of resources or a specific class of operations.

Over the years, several such frameworks have been developed which include the theory of entanglement \cite{Horodecki09, Chitambar_RMP_2019}, coherence \cite{Aberg_arxiv_2006, Plenio_PRL_2014, Winter_PRL_2016, Uttam_PRL_2017, Alexander_RMP_2017}, purity \cite{Horodecki_PRA_2003}, magic \cite{Veitch_NJP_2014, Howard_PRL_2017, Wang_NJP_2019},   asymmetry \cite{Marvian_NJP_2013}, and quantum thermodynamics \cite{Horodecki_PRA_2003, Fernando_PRL_2013, Horodecki13, Fernando_NAS_2015, Renner_PRX_2021}. Each resource theory identifies the resources that serve as the foundation for various quantum information processing. Specifically, 
shared entangled states are  proven to be advantageous for obtaining quantum benefits in classical and quantum information transmission \cite{Bennett_PRL_1992,Bennett_PRL_1993}, as well as in quantum key distribution \cite{Nicolas_RMP_2002}.
 In contrast, all states with non-vanishing purity are the resource in the theory of purity, which is related to quantum thermodynamics  \cite{Horodecki_PRA_2003}. 
Additionally, resource theory provides a method for classifying states. A quantifier is said to be valid if it satisfies certain requirements set forth by a resource theory \cite{Horodecki00}, such as non-negativity, monotonicity under free operations, convexity, etc. Note, interestingly, that the essential prerequisites for a valid measure in a resource theory are strikingly ubiquitous and consistent.

Based on the response to global unitary operations, absolutely separable states (AS) \cite{Marek_PRA_2001, Frank_PRA_2001} which cannot be entangled by any global unitary operations can be separated from  non-absolutely separable states which have no such incapability 
\cite{Kraus_PRA_2001, Henderson_PRA_2003, Chefles_PRA_2005}. 
 Processing quantum information requires identifying such states that can produce an entangled state as an output from separable input states, as is the case in several fields of quantum information science such as quantum tensor network theory \cite{Jacob_arxiv_2019}. There are mathematical criteria based on eigenvalues in spectral decomposition and experimentally-friendly witness operators that can be used to identify non-absolutely separable states \cite{Hilderbrand07, SLATER200917, Johnston13, Arunachalam15, Archan_PRA_2014, Ayan_PRA_2021}. 

In this work, we develop a resource theory of non-absolute separability (NAS). This theory describes the mixture of global unitary operations as free operations that result in absolutely separable states as free states.
We argue that certain desired qualities, such as positivity, monotonicity under free operations, and convexity, should be satisfied by the measures characterizing non-absolute separability. 
Since the set of absolutely separable states is convex and compact, we define distance-based measures to quantify the content of non-absolute separability. Specifically, we prove that independent of the distance measures,  the quantifiers follow all the properties required to be valid non-absolutely separability measures. We demonstrate that  all pure states possess maximal resources for a given dimension.  It is worth noting that the NAS measures are more fine-grained than the entanglement measures, since they are capable of detecting both entangled and a class of non-absolutely separable  states.  Employing relative entropy, Bures and Hilbert-Schmidt metrics, we obtain  exact expressions for NAS measures of pure states.  We also establish a connection between distance-based NAS  and entanglement measures.

Furthermore, we provide  a different class of NAS quantifiers  based on witness operators, which similarly meets all requirements for a legitimate  NAS measure and reduces to a fixed value for all two-qudit pure states in a given dimension. We illustrate and compare the behavior of  all the NAS measures by considering modified Werner states \cite{Werner89}, which is a mixture of a non-maximally entangled state and white noise.  In particular, we discover that the NAS measures for modified Werner states are independent of the parameters in non-maximally entangled states, and they depend solely on the noise strength.

The paper is organized in the following way. In Sec. \ref{sec:resource}, we propose the resource theory for non-absolute separability and its requirements. After discussing the basic conditions
to be satisfied by a NAS measure, we present distance-based measures in Sec. \ref{subsec:distnce} and witness-based measures in Sec. \ref{subsec:witness}. We make concluding remarks in Sec. \ref{sec:conclu}.

\section{Resource Theory for non-absolutely Separable states}
\label{sec:resource}

In order to measure the advantage offered by some resources in  quantum information processing tasks,  
the conceptual framework for quantum resource theory has been developed \cite{Horodecki_IJMPB_2013,Chitambar_RMP_2019}.
Under that structure, quantification and manipulation of quantum resources can be accomplished.
Any resource theory has three fundamentally important characteristics.
\textbf{I.} A class of free operations, $\mathcal{F_O}$ -- resource can not be created by using $\mathcal{F_O}$.
\textbf{II.} A class of free states, $\mathcal{F_S}$ -- these states can be produced via free operations.
 \textbf{III.} Convertibility between resource states under $\mathcal{F_O}$ with an unlimited supply of free states.
  In particular,  it is argued that the transformation between states using free operations should follow a certain rule, which is given the term ``monotones".
It implies that the transformation $\rho \rightarrow \Tilde{\rho}$ under \(\mathcal{F_O}\) is possible with the condition that a specific function, $f$, decreases during the transformation, i.e., $f(\rho)\geq f(\Tilde{\rho})$. More generally, during $\rho \rightarrow \{p_i,~\Tilde{\rho}_i\}$, $f(\rho)\geq \sum_i p_i f(\Tilde{\rho}_i)$, implying that $f$ decreases on average in this picture, thereby obeying the monotonicity under free operations by a measure, $f$ of a given resource theory. Based on this, several resource theories have been proposed in literature \cite{Horodecki09, Chitambar_RMP_2019, Alexander_RMP_2017, Horodecki_PRA_2003, Veitch_NJP_2014, Howard_PRL_2017, Wang_NJP_2019, Marvian_NJP_2013, Horodecki_PRA_2003, Fernando_PRL_2013, Horodecki13, Fernando_NAS_2015, Renner_PRX_2021}. 
Starting from a set of free states, $\mathcal{F_S}$, we  obtain a set of free operations, $\mathcal{F_O}$, by which  the set of free states remain invariant. By taking this route, our aim is to develop a resource theory that is more fine-grained than the theory of entanglement.

\begin{figure}[htb!]
\includegraphics[width=7.0cm, height=5.0cm ]{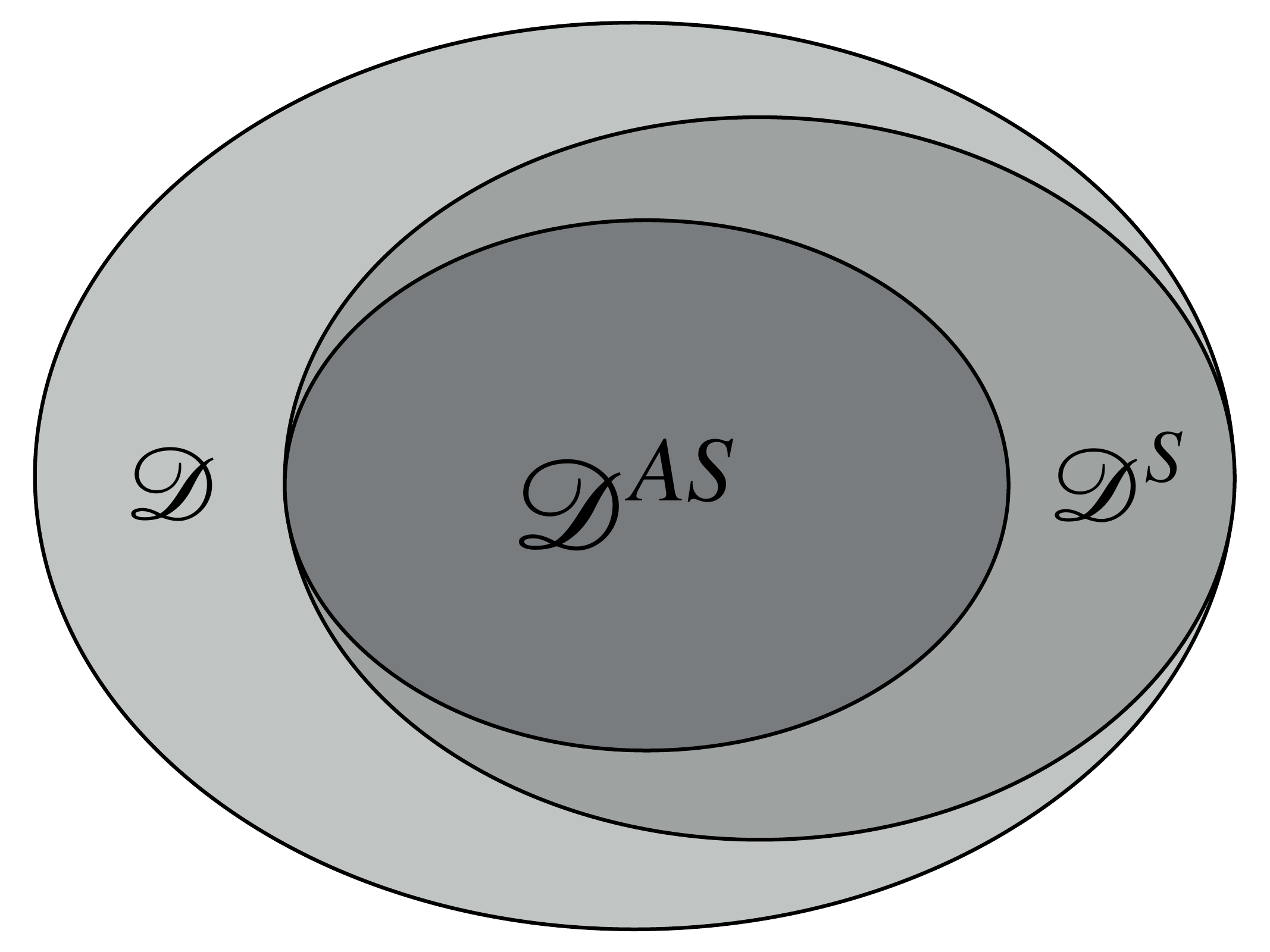}
\caption{A schematic diagram to realize the set of absolutely separable states $\left(\mathscr{D}^{AS}\right)$ and the set of separable states $\left(\mathscr{D}^{S}\right)$ in the set of whole states space $\left(\mathscr{D}\right)$.}
\label{fig:AS_set}
\end{figure}

We are interested in studying a class of separable states that remains separable under any global unitary operations \cite{Kraus_PRA_2001, Henderson_PRA_2003, Chefles_PRA_2005}, known as absolutely separable (AS) states \cite{Marek_PRA_2001, Frank_PRA_2001}. It can be shown that AS states constitute a compact and convex set, $\mathscr{D}^{AS}$ \cite{Archan_PRA_2014}. From the definition, one finds that  under any global unitary operations, an AS state remains AS, otherwise a successive application of several global unitary operations would be able to transform an AS state to an entangled one. Moreover, a mixture of unitary operations,
$$\Lambda_{(p_{i},U_{i})}(\rho)=\sum_{i}p_{i} U_{i}\rho U^{\dagger}_{i},$$
where $\sum_{i}p_{i}=1$ and $0\leq p_i\leq1$, also maps an AS state to an another AS state, since the set $\mathscr{D}^{AS}$ is convex. We refer to any state outside the set $\mathscr{D}^{AS}$ as a non-absolutely separable (non-AS) state. Thus the set of non-AS states contains both entangled and separable states which can be made entangled by some global unitary operations. See Fig. \ref{fig:AS_set} for a pictorial realization.

\textit{Let us now introduce a resource theory of non-absolute separability (NAS), where we define AS states as free states. Therefore, the mixture of global unitary operations $\Lambda_{(p_{i},U_{i})}(\rho)$ can be considered as free operations for this resource theory. Hence, the non-AS states get the status of resourceful states under the NAS resource theory.}

The choice of free operations in quantum resource theory is, in fact, the foundation of its physicality
and is physically reasonable and consistent with the principles of quantum information theory. To argue it,  let us recall that the resource theory of purity \cite{Streltsov_NJP_2018} is based on different types of free operations, namely the mixture of unitaries, $\left(\Lambda^{MU}\right)$, noisy operations $\left(\Lambda^{NO}\right)$, and the unital operations $\left(\Lambda^{U}\right)$. Note that these operations form a subset hierarchy, 
$\Lambda^{MU}\subset \Lambda^{NO} \subset \Lambda^U$. When only the mixture of unitaries is available to us, in that case, the resource theory of purity reduces to the resource theory of NAS. \\

In addition, the NAS resource theory, as we will discuss later, can provide an upper bound of entanglement quantity, and NAS measures are also easily accessible in experiments compared to the entanglement measures. In the succeeding section, we argue that the NAS measure of a state $(\rho)$ depends on its eigenvalue distribution. There are several experimentally feasible techniques to estimate the spectrum of a quantum state \cite{Franconori_spectrum_det,Werner_spectrum_det}, without going through full state tomography. In particular, it is shown how to estimate the spectrum of a quantum state with minimal information in a single experimental setting that can be implemented in linear optical and superconducting systems  \cite{Franconori_spectrum_det}. Moreover, in the case of thermally equilibrium states, one can perform the von-Neumann measurement in the energy eigenbasis to find the spectrum. Using these aforementioned techniques, an upper bound of NAS similar to Eq. (\ref{eq-mixed_bound_relentmeas}) can be evaluated easily, which, in turn, provides an experimental bound on the amount of entanglement, thereby establishing a connection with the resource  theory of entanglement.

Characterization of the AS states, $\rho_{AS}$ defined in $\mathbb{C}^m\otimes\mathbb{C}^n$ can be mapped to the investigation of a set of real numbers $\{\lambda_i\}_{i=1}^{mn}$ such that $\lambda_i\geq\lambda_{i+1}$. 
Every state in that Hilbert space with eigenvalues $\{\lambda_i\}_{i=1}^{mn}$ 
is separable even under the action of any unitary operation $U\in\mathbb{C}^m\otimes\mathbb{C}^n$. In general it is not easy to detect them, although for qubit-qudit states, a criterion to detect AS states is available \cite{Hilderbrand07, SLATER200917, Johnston13, Arunachalam15}. Specifically in a $\mathbb{C}^2 \otimes \mathbb{C}^d$ (in short, denoted as $2\otimes d$)  system, if a bipartite state $\rho_{AB}$ has 2d eigenvalues, $\{\lambda_i^\downarrow\}$ with $\sum_{i=1}^{2d}\lambda_i^\downarrow = 1$, where $\lambda_i^{\downarrow}$ indicates that the elements are ordered in a non-increasing manner, the absolute separability criteria reads
$\lambda_1^{\downarrow} - \lambda_{2d -1}^{\downarrow} -2\sqrt{\lambda_{2d}^{\downarrow}\lambda_{2d-2}^{\downarrow}} \leq 0$. 
\textit{Note that the equality holds for the states lying on the boundary of the set of AS states \cite{Saronath_PRA_2021}}. It is important to note that the set of absolutely positive partially transposed (PPT) states and absolutely separable states coincide in $2\otimes d$ which is not the case in higher dimension \cite{Hilderbrand07}. However, the characterization in higher dimensional Hilbert space is still an open problem that may be studied further. 

Since AS states form a convex and compact set, the Hahn-Banach theorem guarantees a hyperplane that can separate at least one non-AS state from all the AS states. This hyperplane is referred to as a \textit{ witness operator for non-AS states} \cite{Archan_PRA_2014}, which can be an alternative experimentally-friendly method to detect non-AS states. Consequently, we can quantify the non-absolute separability of a non-AS state as the closest distance from the set $\mathscr{D}^{AS}$. 

To characterize non-AS states, we choose here two approaches, namely distance-based and witness-based measures. We will prove that both of them follow  an essential aspect of resource theory, i.e., monotonicity.

Before proceeding further, let us specify the conditions that a NAS measure, $\mathcal{N}(\rho)$ should obey. \\
\textbf{C-I.} (\textit{Positivity}) $\mathcal{N}(\rho) =0$ if and only if $\rho$ is an absolutely separable state or else, $\mathcal{N}(\rho)>0$\\
\textbf{C-II.} (\textit{Invariance under local unitary}) $\mathcal{N}(\rho)$ remains invariant by local unitary operations. $\mathcal{N}(U_{1}\otimes U_{2}\rho U_{1}\otimes U_{2})=\mathcal{N}(\rho)$ \\
\textbf{C-III.} (\textit{Monotonicity}) Under the free operations ($\Lambda$), the measure $\mathcal{N}(\rho)$ cannot increase, i.e., $\mathcal{N}(\Lambda (\rho)) \leq \mathcal{N}(\rho)$.\\
\textbf{C-IV.} (\textit{Convexity}) $\mathcal{N}(\rho)$ is a convex function of the states, i.e., $\mathcal{N}(\sum_i{a_i\rho_i})\leq\sum_i{a_i \mathcal{N}(\rho_i)}$.\\\\
Note that the conditions \textbf{C-I} to \textbf{-III} can be considered as the minimum requirement, thereby necessary conditions, for a valid NAS measure to be satisfied. On the other hand, \textbf{C-IV} can be perceived as  an additional requirement which is typically fulfilled by a ``good" resource quantifier.

\section{Distance-Based Non-absolute separability Measure}
\label{subsec:distnce}

The set of AS states, $\mathscr{D}^{AS}$, is  convex as well as compact  and hence it enables us to obtain a unique minimum distance between a non-AS state and the set $\mathscr{D}^{AS}$. 

\noindent \textbf{Definition 1.} We introduce a new class of measures of the non-absolute separability of a state $\rho$, through the distance measure, $\mathcal{D}(\rho || \rho_{AS})$, as
\begin{equation}
\label{eq-2}
    \mathcal{N}^\chi(\rho) := \underset{\rho_{AS} \in \mathscr{D}^{AS}}{\text{min}} \mathcal{D}^{\chi}(\rho || \rho_{AS} ),
\end{equation}
where $(\mathcal{D}^{\chi})$ is some measure of distance (not necessarily a metric) between two states $\rho$ and $\rho_{AS}$ such that $\mathcal{N}^\chi(\rho)$ satisfies the above three necessary conditions (\textbf{C-I} to \textbf{-III}). $\chi$ represents different distance measures, such as the relative entropy measure $(\mathcal{D}^{R})$, the Bures measure $(\mathcal{D}^{B})$,  the Hilbert-Schmidt measure $(\mathcal{D}^{HS})$ \cite{Vedral_PRL_1997, Vedral_PRA_1998} etc., and the minimization is taken over the set of AS states, $\mathscr{D}^{AS}$. 

Here we enlist the minimal required properties that any distance measure $\mathcal{D}^{\chi}$ should possess to establish it as  a {\it good} resource quantifier.\\
\textbf{P-I.}  $\mathcal{D}^\chi(\rho||\sigma)\geq 0$, where equality holds iff $\rho=\sigma$.\\
\textbf{P-II.} $\mathcal{D}^\chi(\rho||\sigma)$ should be invariant under unitary operations i.e., $\mathcal{D}^\chi(\rho||\sigma) = \mathcal{D}^\chi(U \rho U^{\dagger}||U \sigma U^{\dagger})$.\\
\textbf{P-III.} $\mathcal{D}^\chi(\rho||\sigma)$ should be jointly convex in its both arguments i.e.,
$\mathcal{D}^{\chi}\left(\sum_i a_{i}\rho_{i} || \sum_i a_{i}\sigma_{i}\right) \leq 
    ~ \sum_i a_{i} \mathcal{D}^{\chi}(\rho_i || \sigma_i) $.
\\
\textbf{Remark 0.} It is enough to consider that the distance measure has the property of contraction under a mixture of unitaries, i.e., $\mathcal{D}^\chi(\rho||\sigma)\geq\mathcal{D}^\chi(\Lambda(\rho)||\Lambda(\sigma))$, rather than joint convexity in order to yield a valid measure. However, \textbf{P-III} offers something more that will be explored throughout this Section.

The conditions for a valid NAS measure, \textbf{C-I} and \textbf{C-II}, are directly followed from \textbf{P-I} and \textbf{P-II} respectively, while we will show that \textbf{P-II} and \textbf{P-III} together ensure the validity of \textbf{C-III}.
\\
 Using the joint convexity property of $\mathcal{D}^{\chi}$, we find
\begin{eqnarray}
\label{eq-3}
\mathcal{D}^{\chi}( ~ \Lambda(\rho) || \Lambda(\sigma) ~ ) && = \mathcal{D}^{\chi} \left ( \sum_{i}p_{i} U_{i} \rho U^{\dagger}_{i} ||  \sum_{i}p_{i} U_{i} \sigma U^{\dagger}_{i}               \right ) \nonumber \\ && \leq 
\sum_i p_{i} \mathcal{D}^{\chi}(U_{i} \rho U^{\dagger}_{i} || U_{i} \sigma U^{\dagger}_{i} ) \nonumber \\  && =
\sum_i p_{i} \mathcal{D}^{\chi}(\rho || \sigma ) \nonumber \\  && =
\mathcal{D}^{\chi} (\rho || \sigma  ) \quad \Big(\because \sum_i{p_i}=1\Big),
\end{eqnarray}
where the penultimate line is written by using \textbf{P-II}.\\
Let  $\rho^{*}_{AS}$ be the nearest AS state from $\rho$. Since an AS state remains an AS state under the mixture of global unitary operations ($\Lambda$), $\mathcal{N}^{\chi}(\rho)$ satisfies
\begin{equation} \label{eq-4}
\begin{split}
\mathcal{N}^{\chi}(\rho)=\mathcal{D}^{\chi}(\rho || \rho^{*}_{AS})&  \geq \mathcal{D}^{\chi}\left( \Lambda(\rho) || \Lambda(\rho^{*}_{AS}) \right)  \\
  & \geq  \underset{\rho_{AS} \in \mathscr{D}^{AS}}{\text{min}} \mathcal{D}^{\chi} ( \Lambda(\rho) || \rho_{AS} ) \\
 & = \mathcal{N}^{\chi} (\Lambda(\rho)),
\end{split}
\end{equation}
which implies the monotonicity condition of \textbf{C-III} for all distance-based measures, satisfying \textbf{P-I} to \textbf{P-III}. As a consequence, we obtain that under  free operations of NAS resource theory, the resource cannot rise.
If we now reduce the mixture of global unitary operations to a single global unitary operation ($U$), i.e.,
$\Lambda(\rho) = U \rho U^{\dagger}$,  we have
$
\mathcal{N}^{\chi}(\rho) \geq \mathcal{N}^{\chi}(U \rho U^{\dagger}) = \mathcal{N}^{\chi}(\rho').
$
There exists also a unitary operator $U^{\dagger}$ for which we can write,
$\rho = U^{\dagger} \rho' U = \Lambda'(\rho')$ which leads to
$
\mathcal{N}^{\chi} (\rho') \geq \mathcal{N}^{\chi}(\rho).
$
Therefore, it is only possible if and only if 
 $   \mathcal{N}^{\chi}(\rho) =  \mathcal{N}^{\chi}(U\rho U^\dagger)$.
Hence, the monotonicity of the NAS measure reduces to the equality if the mixture of global unitary operations is replaced by a single global unitary operation. It implies that \textit{the NAS measure of a resource state only depends on the eigenvalues of the state}, i.e., all the states having the same set of eigenvalues are equally resourceful.
\\
\textbf{\textit{Convexity. }}
We are now going to prove an important property, the convexity in its argument, of a {\it good} NAS measure.\\
\textbf{Theorem 1.} Any $\mathcal{N}^{\chi}$ as defined in Definition 1, and equipped with specific distance measure $\chi$ having properties \textbf{P-I} to \textbf{P-III}, is a convex function of the states, i.e.,
\begin{equation}
\label{eq-6}
\mathcal{N}^{\chi}(a_{1}\rho_{1}+ a_{2}\rho_{2}) \leq  a_{1} \mathcal{N}^{\chi}(\rho_{1}) + a_{2} \mathcal{N}^{\chi}(\rho_{2}).
\end{equation}
This means that the mixture of two resource states possesses a less amount of resource than the sum of the individual ones.
\begin{proof}
Let $\rho_{1AS}^\ast$ and $\rho_{2AS}^\ast$ be the nearest AS states from $\rho_1$ and $\rho_2$ respectively. From the joint convexity property of the distance measure \textbf{(P-III)}, it follows
\begin{eqnarray}
\mathcal{N}^{\chi}(a_1\rho_1+a_2\rho_2) && =  \underset{\rho_{AS} \in \mathscr{D}^{AS}}{\text{min}} \mathcal{D}^{\chi}(a_1\rho_1+a_2\rho_2 || \rho_{AS} ) \nonumber\\&&\leq \mathcal{D}^{\chi}(a_1\rho_1+a_2\rho_2 || a_1\rho_{1AS}^\ast+a_2\rho_{2AS}^\ast) \nonumber\\ && \leq a_1 \mathcal{D}^{\chi}(\rho_1||\rho_{1AS}^\ast)+a_2 \mathcal{D}^{\chi}(\rho_2||\rho_{2AS}^\ast) \nonumber \\ && = a_1 \mathcal{N}^{\chi}(\rho_1) + a_2 \mathcal{N}^{\chi}(\rho_2) \nonumber.
\end{eqnarray}
Thus the joint convexity property of a distance measure not only offers a valid but also a {\it good} NAS measure.
\end{proof}

In every resource theory, there exists a maximally resourced state which has the maximum amount of resource, like maximally entangled states in resource theory of entanglement, maximally coherent states in coherence and pure states in purity resource theory etc. Let us identify the maximally resourceful state in the NAS resource theory. \\
$\textbf{Theorem 2.}$ Pure states contain maximal resource in the NAS resource theory.
\begin{proof}
Let us consider an arbitrary density operator $\rho$ acting on $\mathbb{C}^m\otimes\mathbb{C}^n$, which can be written as a convex mixture of pure states, $\ket{\psi_i}\bra{\psi_i}$, i.e., $\rho = \sum_{i} p_{i}\ket{\psi_i}\bra{\psi_i}$.
Since all the states connected by global unitary transformations $\left(\rho\rightarrow U\rho U^\dagger \text{ where } U \text{ is acting on } \mathbb{C}^m\otimes\mathbb{C}^n\right)$ have same resource,  it implies that all the pure states in that Hilbert space should have the same resource. It means, $\mathcal{N}^\chi(\ket{\psi_i})=\mathcal{N}^\chi(\ket{\alpha})$ for any arbitrary pure state $\ket{\alpha}$ in $\mathbb{C}^m\otimes\mathbb{C}^n$. Hence, by using \textit{Theorem 1}, we arrive at  $$\mathcal{N}^\chi(\rho)\leq\mathcal{N}^\chi(\ket{\alpha})~~~\forall~\ket{\alpha}\in\mathbb{C}^m\otimes\mathbb{C}^n,$$
thereby establishing pure states having maximal resource under the NAS resource theory.
\end{proof}
We will now focus on two well-known distance measures, namely the relative entropy and the Bures measures \cite{Wilde}, which satisfy the properties \textbf{P-I} to \textbf{P-III}.

\subsection{Relative entropy of non-absolute separability}

The quantum \textbf{\textit{relative entropy}} between two states, $\rho$ and $\rho_{AS}$ is given by $\mathcal{D}^R(\rho||\rho_{AS})=\text{tr}(\rho\log_2\rho-\rho\log_2\rho_{AS})$. While it is not a metric since it fails to fulfill the property of triangular inequality and is also not symmetric with respect to $\rho$ and $\rho_{AS}$, it satisfies all the required properties as a relevant distance measure for quantifying NAS.
As shown in Ref. \cite{Vedral_PRL_1997}, the relative entropy is a jointly convex function in both of its arguments, and hence $\mathcal{N}^R(\rho)$ also is convex, thereby properly quantifying the NAS. We will now show that for a specific distance measure, we can achieve more.
\\
\textbf{Theorem 3.} For pure states in $2\otimes d$, the relative entropy measure of NAS, $\mathcal{N}^{R}(\rho)=\log_2 \frac{2d+2}{3} $. 
\begin{proof}
 Consider an arbitrary pure state $\ket{\alpha}$ in $2\otimes d$ such that $\rho= \ket{\alpha}\bra{\alpha}$, for which the NAS measure reduces to 
\begin{eqnarray*}
    \mathcal{N}^{R}(\rho) = \underset{\rho_{AS} \in \mathscr{D}^{AS}}{\text{min}} -\text{tr} \{ \rho \log_2 \rho_{AS} \}
\end{eqnarray*}
since for pure states, $S(\rho)=-\text{tr}\{\rho\log_2\rho\}=0$.\\
Let us write $\rho_{AS}$ in spectral decomposition as $\rho_{AS} = \sum_{i=1}^{2d} \lambda_{i}^{\downarrow} \ket{\lambda_{i}} \bra{\lambda_{i}}$, where $\{\lambda_i^\downarrow\}_{i=1}^{2d}$ with the condition $\lambda_i^\downarrow\geq\lambda_{i+1}^\downarrow$ satisfies the absolute separability condition. Using this spectral form, it immediately leads to
\begin{equation*} \label{eq1}
\begin{split}
    \mathcal{N}^{R}(\ket{\alpha}) & = \underset{\rho_{AS} \in \mathscr{D}^{AS}}{\text{min}} -\bra{\alpha}\log_2 \rho_{AS}\ket{\alpha} \\
    & = \underset{\rho_{AS} \in \mathscr{D}^{AS}}{\text{min}} \sum_{i} \log_2 \frac{1}{{\lambda_i}^{\downarrow}} |\bra{{\lambda_{i}}}\ket{\alpha}  |^{2}.
\end{split}
\end{equation*}
Note that $\sum_{i} \log_2 \frac{1}{{\lambda_i}^{\downarrow}} |\bra{\lambda_{i}}\ket{\alpha}  |^{2} \geq \log_2 \frac{1}{{\lambda^{\downarrow}_{1}}}$.

Since $\mathcal{D}^{R}(\rho || \rho_{AS})$ attains its minimum value when $\rho_{AS}$ is on the boundary of the $\mathscr{D}^{AS}$ and the necessary condition for an AS state being on boundary (in $2\otimes d$) is given by \boldsymbol{$\frac{1}{2d} < \lambda^{\downarrow}_{1} \leq \frac{3}{2(d+1)}$} (see Appendix for details), the minimum value that can be achieved by $\log_2\frac{1}{\lambda_1^{\downarrow}}$ in the above inequality is $\log_2 \frac{2d+2}{3}$.
Now construct an AS state $\rho^{*}_{AS}$ as
\begin{equation}
\label{eq-8}
    \rho^{*}_{AS} = \tilde{\lambda}^{\downarrow}_{1}\ket{\lambda_{1}}\bra{\lambda_{1}} + \tilde{\lambda}^{\downarrow}_{2}\ket{\lambda_{2}}\bra{\lambda_{2}} + \dots   + \tilde{\lambda}^{\downarrow}_{2d}\ket{\lambda_{2d}}\bra{\lambda_{2d}},
\end{equation}
where 

\begin{eqnarray}
\label{eq-9}
\tilde{\lambda}^{\downarrow}_{1}= && \frac{3}{2d+2}, \quad \quad  \quad 
\tilde{\lambda}^{\downarrow}_{i}  \geq \tilde{\lambda}^{\downarrow}_{i+1}, \quad \text{and} \; \nonumber \\ &&
\tilde{\lambda}^{\downarrow}_{1}- \tilde{\lambda}^{\downarrow}_{2d-1}-2\sqrt{\tilde{\lambda}^{\downarrow}_{2d}\tilde{\lambda}^{\downarrow}_{2d-2}} = 0 ,
\end{eqnarray}
 
such that 
\begin{equation}
\label{eq-10}
  \ket{\lambda_{1}} = \ket{\alpha}, \quad  \quad \braket{\lambda_i}{\alpha}_{i\neq1} =0.  
\end{equation}

For this AS separable state $\left(\rho^*_{AS}\right)$,
\begin{equation}
    \label{eq-11}
    \begin{split}
       \sum_{i} \log_2 \frac{1}{{\tilde{\lambda}_i}^{\downarrow}} |\bra{\lambda_{i}}\ket{\alpha}  |^{2} \mathbf{\Bigg{|}_{min}} & = \log_2\frac{2d+2}{3}.
    \end{split}
\end{equation}
Therefore, we can always construct an AS state $\rho^{*}_{AS}$ for which the minimum distance from an arbitrary pure state $\ket{\alpha}$ becomes ${\log_2 \frac{2d+2}{3}}$ and hence for any arbitrary pure state, the relative entropy measure of NAS is ${\log_2 \frac{2d+2}{3}}$.
\end{proof}

\textbf{Example.} Consider an arbitrary pure state $\ket{\alpha}$ in $2\otimes 2$. We can build an AS state by mixing four mutually orthogonal pure states $\{\ket{\alpha},\ket{\alpha^{\perp}},\ket{\alpha^{\perp\perp}},\ket{\alpha^{\perp\perp\perp}}\}$ as
\begin{eqnarray}
\label{eq-2*2nearestAS}
   \rho^*_{AS}=&& \frac{1}{2}\ket{\alpha}\bra{\alpha} + \frac{1}{6}\Big[\ket{\alpha^{\perp}}\bra{\alpha^{\perp}} + \ket{\alpha^{\perp\perp}}\bra{\alpha^{\perp\perp}}  \nonumber\\&&+\ket{\alpha^{\perp\perp\perp}}\bra{\alpha^{\perp\perp\perp}}\Big] \nonumber \\ = &&  
   \frac{1}{3} \ket{\alpha}\bra{\alpha} + \frac{2}{3} \frac{\mathds{I}_{4\times4}}{4},
\end{eqnarray}
where $\ket{\alpha^{\perp}}, \ket{\alpha^{\perp\perp}},$ and $\ket{\alpha^{\perp\perp\perp}}$ can be any three mutually orthogonal vectors in $2\otimes2$ that are also orthogonal to $\ket{\alpha}$.
It is evident that $\rho^*_{AS}$ lies on the boundary of $\mathscr{D}^{AS}$ in $2\otimes2$ which leads to $\mathcal{N}^{R}(\ket{\alpha})=1$.
\\\\
\textbf{Remark 1.} \textit{ From Theorems 2 and 3, it follows that in $2\otimes d$, the maximum relative entropy measure of NAS, $\mathcal{N}^R(\rho)$ can be, $\log_2\frac{2d+2}{3}$}.

\subsection{Non-absolute separability via the Bures distance}

The \textbf{\textit{Bures distance}} between two states $\rho$ and $\rho_{AS}$, $d_{B}(\rho, \rho_{AS}) = \sqrt{2 -2 \sqrt{F(\rho,\rho_{AS})}} $ \cite{Hubner_PLA_1992}, is a valid metric where $F(\rho, \rho_{AS})$ is the Uhlmann fidelity, 
$F(\rho, \rho_{AS})= [ \text{tr}\{\sqrt{\sqrt{\rho} \rho_{AS} \sqrt{\rho}}\}]^{2}$ \cite{Uhlmann76, Jozsa94}. Like the measure of entanglement, we take $\mathcal{D}^{B}(\rho || \rho_{AS})=d^{2}_B(\rho, \rho_{AS})= 2 -2 \sqrt{F(\rho,\rho_{AS})}$ \cite{Vedral_PRL_1997, Vedral_PRA_1998} as the NAS measure. Note that, although $\mathcal{D}^{B}(\rho || \rho_{AS})$ is not a valid metric, it still satisfies all the required properties i.e., \textbf{P-I} to \textbf{III}. In particular, since the Bures metric is a true metric, the corresponding Bures measure,  $\mathcal{D}^{B}(\rho || \rho_{AS})$,
satisfies \textbf{P-I}, and \textbf{P-II} is obvious from its definition, and since $\sqrt{F(\rho,\rho_{AS})}$ is a jointly concave function in both of its arguments \cite{Nielsen_Chuang},  $\mathcal{D}^{B}(\rho || \rho_{AS})$ also enjoys the jointly convex property, i.e., \textbf{P-III}.\\
Like relative entropy, the Bures measure of NAS reduces to a compact form for pure states.
\\
\textbf{Theorem 4.}  For pure states in $2\otimes d$, the Bures measure of NAS, $\mathcal{N}^{B}(\rho)= 2-2\; \sqrt{\frac{3}{2d+2}} $.
\begin{proof}
 For an arbitrary pure state $\rho=\ket{\alpha}\bra{\alpha}$,
\begin{eqnarray}
    \mathcal{N}^{B}(\rho) =&&\underset{\rho_{AS} \in \mathscr{D}^{AS}}{\text{min}}  \Big[2 -2\; \text{tr} \Bigl\{ \sqrt{ \sqrt{\rho} \rho_{AS} \sqrt{\rho}} \; \Big \}\Big]   \nonumber\\ = &&
    2 -2\underset{\rho_{AS} \in \mathscr{D}^{AS}}{\text{max}} \sqrt{\bra{\alpha}\rho_{AS}\ket{\alpha}    } \nonumber\\ = &&
    2 -2\underset{\rho_{AS} \in \mathscr{D}^{AS}}{\text{max}}\sqrt{\sum_i\lambda_i^\downarrow|\braket{\alpha}{\lambda_i}|^2}, \nonumber
\end{eqnarray}
where, in second line, we use the fact that $\sqrt{\rho}=\rho$ as $\rho$ is pure and the last line is obtained by using the spectral decomposition of $\rho_{AS}$. 
Moreover, we again have $\sum_{i}\lambda^{\downarrow}_{i} |\bra{\alpha}\ket{\lambda_{i}} |^{2} \leq \lambda^{\downarrow}_{1}$.

As we argued earlier, the minimum value of $\mathcal{D}^{B}(\rho || \rho_{AS})$ can be achieved when $\rho_{AS}$ lies on the boundary of $\mathscr{D}^{AS}$, and the constraint for an AS state being on the boundary is $\frac{1}{2d} < \lambda^{\downarrow}_{1} \leq \frac{3}{2(d+1)}$. By choosing  $\rho^{*}_{AS}$ as in Eq. (\ref{eq-8}) with all the  conditions followed by Eq. (\ref{eq-9}-\ref{eq-10}), we have
\begin{equation}
\label{eq-bures_measure_for_pure}
    \mathcal{N}^{B}(\ket{\alpha}) = 2-2\; \sqrt{\frac{3}{2d+2}}.
\end{equation}
Hence proved.
\end{proof}
\textbf{Remark 2.} \textit{ From Theorems 2 and 4, it can be seen that the maximum value of $\mathcal{N}^B(\rho)$ in $2\otimes d$ is $2-2\; \sqrt{\frac{3}{2d+2}}$}.

\textbf{Example.} To illustrate our results, let us consider a modified Werner state in $2\otimes2$ \cite{Werner89},
\begin{equation}
\label{eq-modified_werner_state}
    \rho_{W} = p \ket{\xi}\bra{\xi} + \frac{1-p}{4}  \mathds{I}_{4\times 4},
\end{equation}
where $\ket{\xi} = \cos\gamma\ket{00} + e^{i\phi} \sin\gamma \ket{11}$, with \(0\leq \gamma \leq \pi\) and \(0\leq \phi \leq 2 \pi\).

We find that  the state is AS in $0 \leq p \leq \frac{1}{3} $ and it is entangled in $\frac{1}{1+2\sin 2\gamma} < p \leq 1 $, i.e., when $\frac{1}{3} < p  \leq \frac{1}{1+2\sin 2\gamma} $, the state is separable but not AS. However, the state is non-AS in the range $\frac{1}{3}< p\leq1$.

Let us compute the NAS measure for the state $\rho_W$ with both relative entropy and the Bures measure as the distance.

\textbf{Proposition 1.}
The NAS content of the modified Werner state ($\rho_W$), in the range $\frac{1}{3} < p\leq1$, reads as
\begin{eqnarray}
    \mathcal{N}^{R}(\rho_W) = && \log_2 6 -\frac{1+3p}{4} \log_2 3 -S(\rho), \,\, \mbox{and} \nonumber\\
      \mathcal{N}^{B}(\rho_W)  = &&
    2 - \left[ \sqrt{\frac{1+3p}{2}} + \sqrt{\frac{3}{2}(1-p)  } \right], \nonumber
\end{eqnarray}
with respect to the relative entropy and Bures measures of NAS respectively.
\begin{proof}
 For a mixed state $\rho$ defined on $2\otimes d$,  we can write
\begin{eqnarray}
    \mathcal{N}^{R}(\rho)  = 
    \underset{\rho_{AS} \in \mathscr{D}^{AS}}{\text{min}} \left[- S(\rho) - \text{tr} (\rho \log_2 \rho_{AS})\right].\nonumber
\end{eqnarray}
We write $\rho_{AS}$ and $\rho$  in spectral forms as
\begin{equation}
\label{eq-spectral_form}
\rho_{AS} = \sum_{i=1}^{2d} \lambda_{i}^{\downarrow} \ket{\lambda_{i}} \bra{\lambda_{i}},~~~~
\rho = \sum_{j=1}^{2d} \alpha_{j}^{\downarrow} \ket{\alpha_{j}} \bra{\alpha_{j}},
\end{equation}
where $\alpha^{\downarrow}_{i} \geq \alpha^{\downarrow}_{i+1}$ and same for $\lambda^{\downarrow}_{i}$.
Hence
\begin{eqnarray}
\label{eq-relent_meas_gen}
\mathcal{N}^{R}(\rho) && = \underset{\rho_{AS} \in \mathscr{D}^{AS}}{\text{min}} \left[ \sum_{i}  \log_2 \frac{1}{\lambda^{\downarrow}_i} \bra{\lambda_i}\rho \ket{\lambda_i}  - S(\rho)\right] \nonumber \\  && =
\underset{\rho_{AS} \in \mathscr{D}^{AS}}{\text{min}} \left[ \sum_{i}  p_i \; \log_2 \frac{1}{\lambda^{\downarrow}_i}  - S(\rho) \right ],
\end{eqnarray}
where $ p_i = \bra{\lambda_i}\rho \ket{\lambda_i})$ are the probabilities, i.e., $0\leq p_i\leq1$ and $\sum_i p_i=1$. Considering $\ket{\lambda_i}=\sum_jc_j^i\ket{\alpha_j}$, we have $p_i=\sum_j\alpha_j^\downarrow\big|c_j^i\big|^2$ with the condition $\sum_j\left(c_j^{k}\right)^*c_j^i=\delta_{ik}$.\\
Since the eigenvalues of $\rho_{AS}$ are arranged in non-increasing order, so $\log_2 (\frac{1}{\lambda^{\downarrow}_{i}}) \leq \log_2 (\frac{1}{\lambda^{\downarrow}_{i+1}})$, hence to get the minimum of $\mathcal{N}^{R}(\rho)$, the probabilities $p_i$ should satisfy the condition $p_i \geq p_{i+1}$ \cite{Hilderbrand07} with the inequality constraint $\alpha_{2d}^\downarrow\leq p_i\leq\alpha_1^\downarrow$.  In other words, we have to minimize the above expression (Eq. (\ref{eq-relent_meas_gen})) with respect to $p_{i}$$^{,}$s (or $c_j^i$$^{,}$s) with the aforementioned conditions and $\lambda_i^\downarrow$$^{,}$s with the constraint specified in Eq. ({\ref{eq-9}}).\\\\
For the modified Werner state, $\rho_W$, we find that the minimum is achieved when $\lambda_1^\downarrow=\frac{1}{2}$, $\lambda_{i=2,3,4}^\downarrow=\frac{1}{6}$ and $p_i=\alpha_i^\downarrow$ (i.e., $\ket{\lambda_i}=\ket{\alpha_i}$). Note that, for the state $\rho_W$, $\alpha^{\downarrow}_1 = \frac{1+3p}{4}$ , $\alpha^{\downarrow}_{i=2,3,4} = \frac{1-p}{4}$. Hence the relative entropy measure of NAS for the state $\rho_W$ is found to be 

\begin{equation}
\label{eq-NASmeasure_werner_relent}
    \mathcal{N}^{R}(\rho_W) = \log_2 6 -\frac{1+3p}{4} \log_2 3 -S(\rho).
\end{equation}

For the Bures measure of NAS, it reads as

\begin{eqnarray}
\label{eq-NASmeasure_werner_bures}
    \mathcal{N}^{B}(\rho_W) &&= 2 - 2 \sum_{k=1}^4 \sqrt{\alpha^{\downarrow}_{k} \lambda^{\downarrow}_{k}} \nonumber \\ &&= 
    2 - \left[ \sqrt{\frac{1+3p}{2}} + \sqrt{\frac{3}{2}(1-p)  }      \right].
\end{eqnarray}
\end{proof}
The above exercise has tempted us to conjecture that for any state $\rho$ in $2\otimes d$, the minimization in Eq. (\ref{eq-relent_meas_gen}) over $p_i$$^{,}$s can be achieved if $p_i=\alpha_i^\downarrow$,   i.e., $\ket{\lambda_i}_{i=1 \dots 2d} = \ket{\alpha_i}_{i=1 \dots 2d}$. Consider a state $\Tilde{\rho}_{AS}$ lying on the boundary of $\mathscr{D}^{AS}$, has eigenvalues $\lambda_1^\downarrow=\frac{3}{2d+2}$ and $\lambda_{i=2,3,\dots,2d}^\downarrow= \frac{1}{2d+2}$ with $\ket{\lambda_i}=\ket{\alpha_i}$.\\
\textit{Upper bound on the distance-based measures of NAS for any arbitrary state $\rho$ in $2\otimes d$. }   The relative entropy between $\rho$ and $\Tilde{\rho}_{AS}$ which is chosen via the above prescription reduces to
\begin{eqnarray}
\label{eq-mixed_bound_relentmeas}
    \mathcal{N}^{R}(\rho) \leq && \alpha_1^\downarrow \log_2 \frac{2d+2}{3} + (1-\alpha_1^\downarrow) \log_2 (2d+2) -S(\rho) \nonumber \\ \leq &&
    \log_2 (2d+2) - \alpha_1^\downarrow \log_2 3  - S(\rho),
\end{eqnarray}

while in case of \textit{the Bures measure of NAS}, it reads as 
\begin{eqnarray}
\label{eq-mixed_bound_buresmeas}
    \mathcal{N}^{B}(\rho) \leq 2-2\left[ \frac{\text{tr}\sqrt{\rho}+(\sqrt{3}-1)\sqrt{\alpha_1^\downarrow}}{\sqrt{2d+2}}\right].
\end{eqnarray}

\begin{figure}[htb!]
\includegraphics[width=\linewidth ]{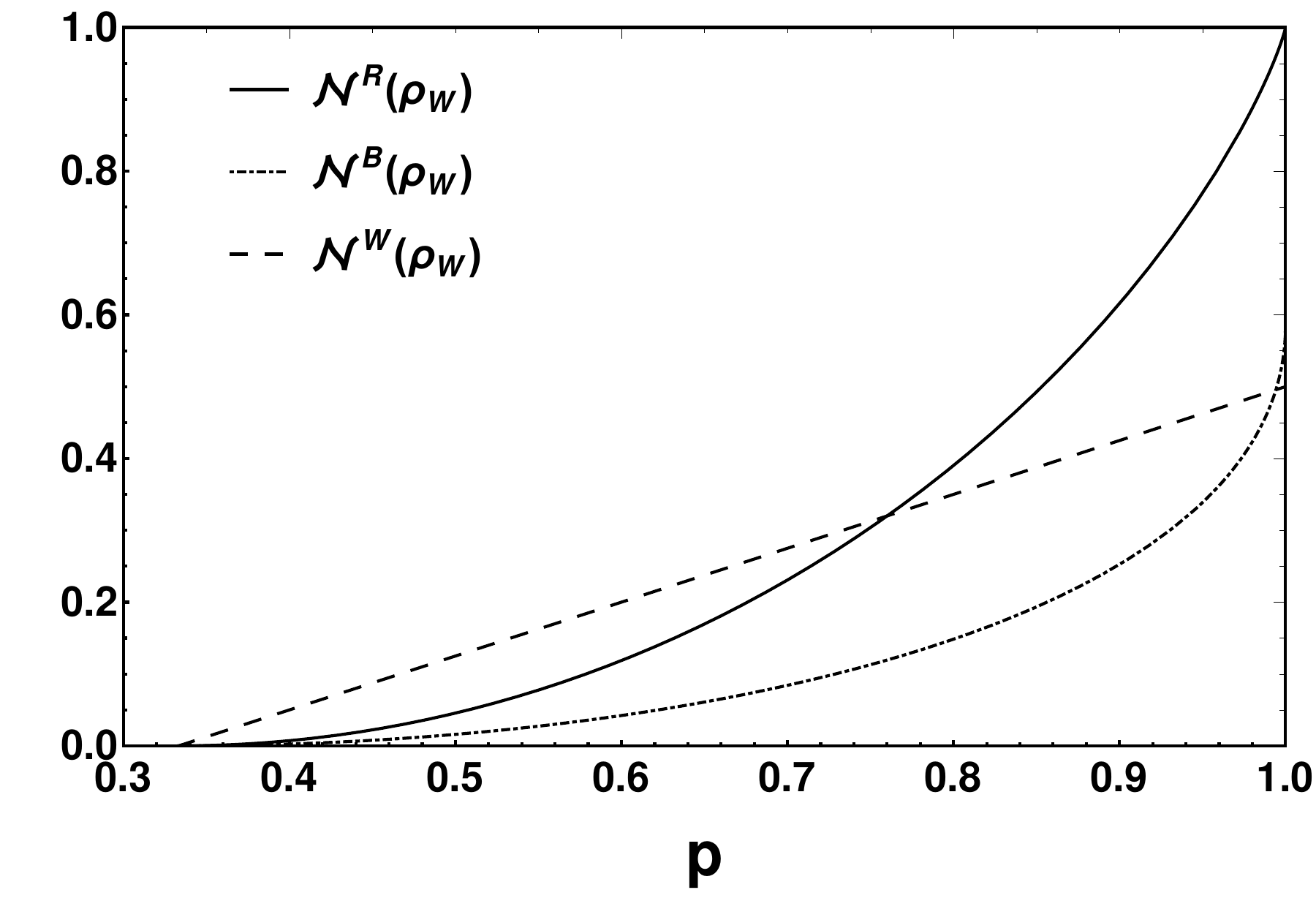}
\caption{ Relative entropy measure $(\mathcal{N}^{R}(\rho_{W}))$ (solid  line), Bures measure $(\mathcal{N}^{B} (\rho_{W}))$ (dot-dashed line) and witness-based measures $(\mathcal{N}^{W} (\rho_{W}) )$ (dashed line) for NAS of the modified Werner state $\rho_W$ with respect to $p$ (abscissa). Note here that  
the NAS measure is independent of \(\gamma\) and \(\phi\) involved in non-maximally entangled state. Note further that $p=1$ corresponds to a pure state for which $\mathcal{N}^{R}(\rho_W) =1$,  $\mathcal{N}^{B}(\rho_W) =2-\sqrt{2}$ and $\mathcal{N}^{W}(\rho_W) =1/2$. All the axes are dimensionless. }
\label{fig:werner_quant}
\end{figure}

Note that the inequality for pure states indeed reduces to the equality as already shown in \textit{Theorems 3} and \textit{4}.\\
Fig. \ref{fig:werner_quant} depicts $\mathcal{N}^{R}(\rho_W)$ and $\mathcal{N}^{B} (\rho_W)$ of the modified Werner state $\rho_W$ 
with respect to $p$. 
Note that the NAS measure $\mathcal{N}^\chi(\rho_W)$ does not depend on \(\gamma\) and \(\phi\), it depends  only on the state parameter $p$ since the eigenvalues of \(\rho_{W}\) are functions of \(p\) as proved in \textit{Proposition 1}. We can see that both of the NAS measures increase with respect to $p$ and reach its maximum value at $p=1$, i.e., when $\rho_W$ is a pure state.

\subsection{Metric-based non-absolute separability measure}
Let us propose an NAS measure, based on a \emph{true metric}, following additional characteristics such as symmetry under its arguments and triangular inequality, which are not obeyed by relative entropy and Bures measures. Further, the metrics, denoted by $\mathcal{D}^\Xi(\rho||\rho_{AS})$, are jointly convex.
\\
\textbf{Theorem 5.} 
Under a jointly convex metric $\mathcal{D}^\Xi(\rho||\rho_{AS})$, if  $\rho^{*}_{AS}$ is the nearest AS state from $\rho$, it remains so for all those states which can be written as convex linear combination of these two states $\rho$ and $\rho^{*}_{AS}$.

\begin{proof}
 Let $\rho^{*}_{AS}$ be the nearest AS state from $\rho$. Now consider a state $\rho_{x}=x\rho + (1-x)\rho^{*}_{AS}$, where $0\leq x\leq1$.
By using the joint convexity of $\left (\mathcal{D}^{\Xi}(\rho || \sigma)\right )$ and the properties of a metric, we get
\begin{eqnarray*}
\label{eq-rhorhox}
\mathcal{D}^\Xi(\rho || \rho_x) && = \mathcal{D}^\Xi(\rho || x\rho + (1-x)\rho^{*}_{AS}) \nonumber \\ &&
\leq (1-x) \mathcal{D}^\Xi(\rho || \rho^{*}_{AS}), ~~~\text{and}\nonumber\\
\mathcal{D}^\Xi(\rho^{*}_{AS} && || \rho_x) \leq x \; \mathcal{D}^\Xi(\rho^{*}_{AS} || \rho),
\end{eqnarray*}
which leads to $\mathcal{D}^\Xi(\rho || \rho_x) + \mathcal{D}^\Xi(\rho_x || \rho^{*}_{AS}) \leq \mathcal{D}^\Xi(\rho || \rho^{*}_{AS})$,
by using $\mathcal{D}^{\Xi}(\rho_{AS}^* || \rho_x)=\mathcal{D}^{\Xi}(\rho_x || \rho_{AS}^*)$.\\
On the other hand, from the triangular inequality, we can write $\mathcal{D}^\Xi(\rho || \rho_x) + \mathcal{D}^\Xi(\rho_x || \rho^{*}_{AS}) \geq \mathcal{D}^\Xi(\rho || \rho^{*}_{AS})$.
Combining these, we obtain
\begin{equation}
\label{eq-equality}
   \mathcal{D}^\Xi(\rho || \rho^{*}_{AS}) = \mathcal{D}^\Xi(\rho || \rho_x) + \mathcal{D}^\Xi(\rho_x || \rho^{*}_{AS}).
\end{equation}
To prove that $\rho^{*}_{AS}$ is also the nearest AS state for $\rho_x$, consider an arbitrary AS state $\rho_{AS}$ for which we can write
\begin{eqnarray*}
    && \mathcal{D}^\Xi(\rho_x || \rho_{AS}) - \mathcal{D}^\Xi(\rho_x || \rho^{*}_{AS}) \nonumber \\ =&&
    \mathcal{D}^\Xi(\rho_{x}||\rho_{AS}) + \mathcal{D}^\Xi(\rho_{x}||\rho) - \mathcal{D}^\Xi(\rho_{x}||\rho^{*}_{AS}) - \mathcal{D}^\Xi(\rho_{x}||\rho) \nonumber \\ \geq &&
    \mathcal{D}^\Xi(\rho||\rho_{AS}) - \mathcal{D}^\Xi(\rho||\rho^{*}_{AS}) \geq 0.
\end{eqnarray*}
Hence, we can claim that $\rho^{*}_{AS}$ is also the nearest AS state for $\rho_x$.
\end{proof}
Let us move to a family of metrics that have some special properties.
Consider two quantum states $\rho$ and $\rho_{AS}$ for which we can define a family of metrics as \cite{Nicolas_1986}
\begin{equation*}
\label{eq-family_of_metrics}
d_{p}(\rho,\rho_{AS}) = \underset{\{P_j\}_{j=1}^K}{\sup} \left( \sum_{j=1}^{K} \Big| (\tr \rho P_j)^{\frac{1}{p}} - (\tr \rho_{AS}P_{j})^{\frac{1}{p}} \Big |^{p}  \right )^{\frac{1}{p}},
\end{equation*}
where $p$ is a fixed positive integer and the supremum is taken over all finite families of projectors $\{P_j\}_{j=1}^K$ such that $\sum_{j=1}^{K} P_j =\mathds{I}$. Note that for $p=1$, $ d_{1}(\rho, \rho_{AS}) = \text{tr} |\rho - \rho_{AS} |$, while with $p=2$, we recover the Bures metric, i.e., $d_{2}(\rho, \rho_{AS})=d_B(\rho, \rho_{AS})$.
In Ref. \cite{Ma_2011}, $d_{1}(\rho, \rho_{AS})$ is shown to be a jointly convex and contractive metric under a completely positive and trace preserving (CPTP) map, and hence it can be used to quantify both the entanglement and the non-absolute separability. 
\\
\textbf{Corollary.} The distance measures which are jointly convex, obey all the metric properties, and the contractive metric under a CPTP map leads to NAS and entanglement measures such that an upper bound on entanglement measure can be given by
$$ \mathcal{N}^\Xi(\rho) - \mathcal{N}^\Xi(\rho_{x^{*}}) \geq E(\rho),$$ where $\mathcal{N}^\Xi(\rho)$ and $E(\rho)$ are the distance-based NAS and entanglement measures \cite{Vedral_PRL_1997} of $\rho$. Here $\rho_{x^{*}}$ is the nearest separable state from $\rho$ of the form 
$
\rho_{x^{*}} = x^{*} \rho + (1-x^{*})\rho^{*}_{AS}.
$
\begin{proof}
Let $\rho^{*}_{AS}$ be the nearest AS state from an entangled state $\rho$, and let a state $\rho_{x}$ be
 $   \rho_{x}=x\rho + (1-x)\rho^{*}_{AS}$.
 From Eq. (\ref{eq-equality}), we can write
\begin{eqnarray*}
     \mathcal{D}^{\Xi}(\rho || \rho_{x})  = && \mathcal{D}^{\Xi} (\rho || \rho^{*}_{AS}) -  \mathcal{D}^{\Xi} (\rho_{x} || \rho^{*}_{AS}) \nonumber \\ = &&
     \mathcal{N}^{\Xi}(\rho) - \mathcal{N}^{\Xi}(\rho_x),
\end{eqnarray*}
where the last line follows from the \textit{Theorem 5} and the \textit{Definition 1}.

The distance-based \textit{entanglement measure}  can be defined as \cite{Vedral_PRL_1997}
 $  E(\rho) = \underset{\rho_{S} \in \mathscr{D}^{S}}{\text{min}} \mathcal{D}^{\Xi} (\rho || \rho_{S})$,
where $\mathscr{D}^S$ is the set of separable states.
If $\rho_{x}$ is a separable state,  $E(\rho) \leq \mathcal{D}^\Xi(\rho || \rho_{x})$.
 
 The bound can be made tighter if $\rho_x$ lies on the boundary of $\mathscr{D}^S$. We consider $\rho_{x^{*}}$ to be such a state,
 which is the nearest separable state from $\rho$ of the form $
\rho_{x^{*}} = x^{*} \rho + (1-x^{*})\rho^{*}_{AS}
$. Hence, we obtain a tighter upper bound on the entanglement content of $\rho$ as
\begin{eqnarray}
\label{eq-tighter_upperbound}
     E(\rho)  \leq && D^\Xi(\rho || \rho_{x^{*}}) \nonumber \\ = &&
     \mathcal{N}^\Xi(\rho) - \mathcal{N}^\Xi (\rho_{x^{*}}).
\end{eqnarray}
Boundary of $\mathscr{D}^S$ can be characterized by all those states which cannot be written as a convex linear combination of a maximally mixed state and another separable state, i.e.,
$$
\rho_{x^{*}} \neq y \; \frac{\mathds{I}}{d} + (1-y) \rho_{S}~~~ \forall ~ \rho_{S}\in \mathscr{D}^S
$$
and for any $0 \leq y \leq 1 $.
\end{proof}
\textbf{Example.}  Let us take a pure  state in 2 $\otimes$ 2, $\ket{\psi} = \beta \ket{00} + \delta \ket{11}$ with $\{\beta,\delta\}\in[0,1]$. We can obtain this form by computing Schmidt decomposition of an arbitrary pure state in 2 $\otimes$ 2 and then transforming it via a local unitary transformation. Note that both the measure of entanglement and non-absolute separability are invariant under a local unitary transformation. The nearest separable state of $\ket{\psi}$ is given by $\beta^2 \ket{00}\bra{00} + \delta^2 \ket{11}\bra{11}$ \cite{Vedral_PRA_1998}, whereas the nearest AS state is $\rho^{*}_{AS} = \frac{1}{3} \ket{\psi}\bra{\psi}+ \frac{2}{3} \frac{\mathds{I}_{4\times 4}}{4} $ (see Eq. (\ref{eq-2*2nearestAS})). So, we can write $\rho_x=x\ket{\psi}\bra{\psi}+(1-x)\rho_{AS}^*=p\ket{\psi}\bra{\psi}+(1-p) \frac{\mathds{I}_{4\times 4}}{4}$ with $p=\frac{2x+1}{3}$. Using Peres-Horodecki criteria \cite{Peres96, horodecki96a} for separability in $2\otimes2$, we check that $\rho_x$ is separable in the range $0\leq x\leq \frac{1}{2}\left(\frac{3}{1+4\beta\delta}-1\right)$. Hence, we can maximize $\mathcal{D}^\Xi(\rho_x||\rho_{AS}^*)$ over $x$ in the aforementioned range in order to find $\mathcal{N}^\Xi (\rho_{x^{*}})$.

\begin{figure}[htb!]
\includegraphics[width=\linewidth ]{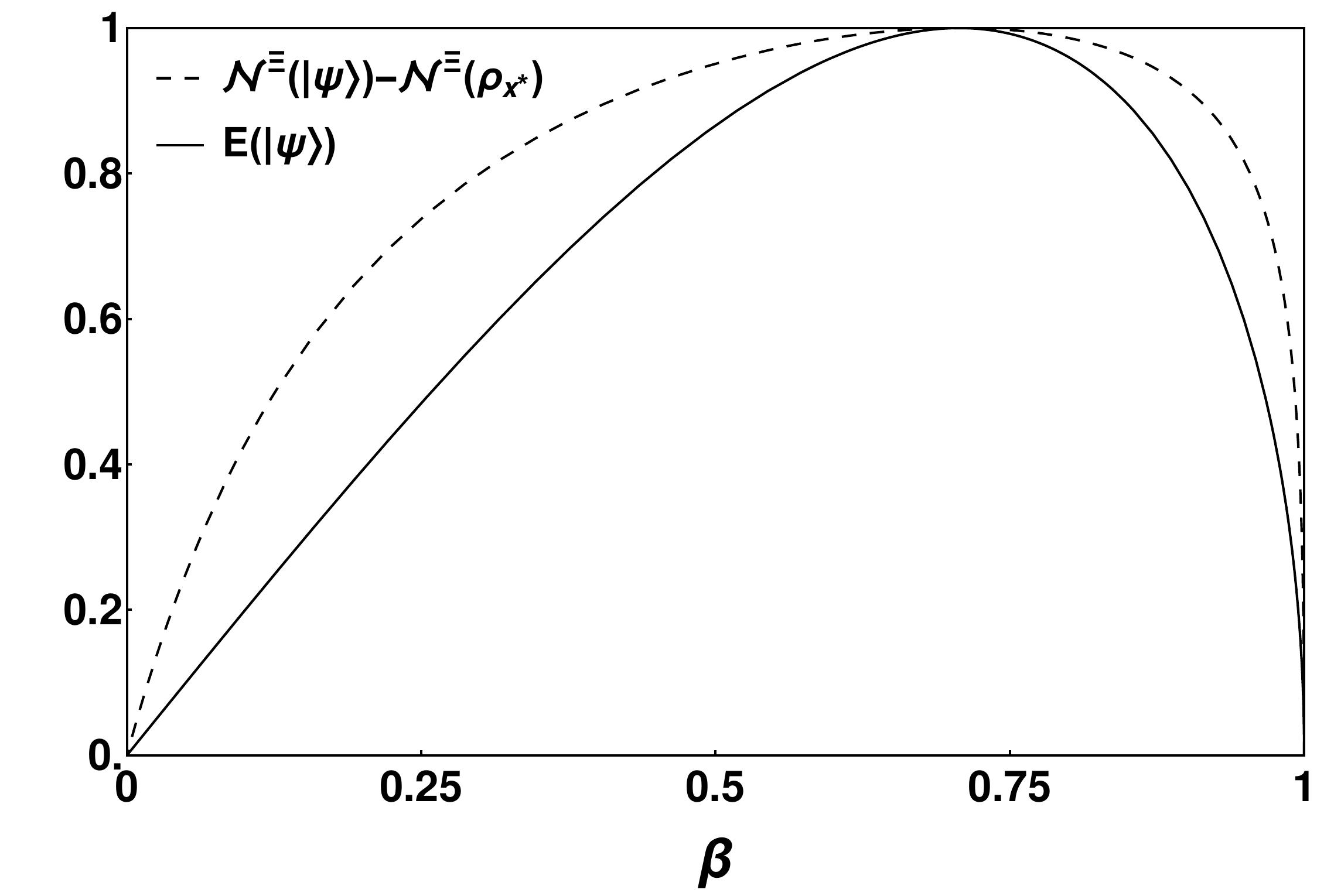}
\caption{The entanglement measure $(E(\rho))$ (solid line) and the upper bound of the entanglement $(\mathcal{N}^\Xi(\rho) - \mathcal{N}^\Xi (\rho_{x^{*}}) )$ (dashed line) of the state $\ket{\psi}$ with respect to $\beta$ (abscissa). Note that at $\beta=\frac{1}{\sqrt{2}}$, the state is maximally entangled and the upper bound of entanglement coincides with its exact value. All the axes are dimensionless.      }
\label{fig:bound_ent}
\end{figure}

Fig. \ref{fig:bound_ent} shows the entanglement content of the state $\ket{\psi}$ and its upper bound as computed by our distance-based NAS measure with the 
 metric $d_1(\rho, \sigma)$. Note that the upper bound of entanglement coincides with its exact value for the maximally entangled state.

\section{Characterization of non-absolutely separable states via witness operators}
\label{subsec:witness}

We now move to prescribe another measure to quantify NAS which is based on the witness operators. In $d_1\otimes d_2$ dimension with $d_1, d_2 \geq3$, the characterization of AS states is  hard
although non-AS states can be detected by using witness operators \cite{Archan_PRA_2014, Ayan_PRA_2021}, which are  useful method to identify resourceful states  in experiments.
It was shown that entanglement measures quantifying resources can be obtained from the witness operators  \cite{Fernando_PRA_2005}. 

Following this idea, we propose here a NAS measure based on witness operators. 
Since AS states constitute a convex and compact set which enables the construction of a hyperplane to separate all the AS states from at least one non-AS state,  this hyperplane can witness that non-AS state. Consider a non-AS state, $\rho$ which may be separable or entangled. It implies that  there exist some unitary operators $U$ which can turn this state (if separable) into an entangled one or can increase its (if already entangled) entanglement content.  Anyway, one can detect this   state as a non-AS state through an entanglement witness operator $W$ after the operation of global unitary, $U$. This is basically the standard way to detect a non-AS state \cite{Archan_PRA_2014}.

A witness operator $(\tilde{W})$ for detecting a non-AS state is defined as $\text{tr}\{ \tilde{W} \rho_{AS}\}\geq 0 ~~\forall~~\rho_{AS}\in \mathscr{D}^{AS}$, and $\text{tr} \{\tilde{W}\rho\}<0$ for at least one non-AS state $\rho$, with

$\tilde{W}$ as $U^\dagger W U$, where $U$ is a global unitary operator and $W$ is an entanglement witness operator.
Equipped with these notions, one can define an NAS resource monotone called, \textit{witness-based NAS measure}, grounded on witness operator $(\tilde{W})$. \\
\textbf{Definition 2.} The \textit{witness-based quantifier} ($\mathcal{N}^{W}(\rho)$) of a non-AS state $\rho$ is defined as
\begin{eqnarray}
\label{eq-witness_measure}
    \mathcal{N}^{W}(\rho) =&& \text{max} \left [ 0,~ -   \underset{\{W \in \mathcal{M};~U\}}{\text{min}} \text{tr}\{U^\dagger W U \rho \}   \right ]\nonumber\\
    =&&\text{max}\left[0,~-\text{tr}\{U_\rho^\dagger W_\rho U_\rho \rho)\}\right]\nonumber\\
  =&&\text{max}\left[0,~-\text{tr}\{ \tilde{W}_\rho \rho)\}\right],
\end{eqnarray}
where $\mathcal{M}= W\cap\mathcal{C}$ is the intersection of the set of entanglement witnesses $(W)$ with some other set $(\mathcal{C})$ such that $\mathcal{M}$ is compact \cite{Fernando_PRA_2005}. In Eq. (\ref{eq-witness_measure}), $\tilde{W}_\rho$ stands for the optimal non-AS witness operator for the state $\rho$.\\
The witness-based NAS measure indeed satisfies all the necessary conditions mentioned as \textbf{C-I} to \textbf{C-III} to be a valid NAS measure. To verify this, consider the following. \textbf{C-I} is directly followed from the definition of the witness operator, and the invariance of the NAS measure under the global unitary operation is evident from the definition of the witness-based NAS measure itself, which validates \textbf{C-II} as well. Now, we prove the monotonicity condition i.e., \textbf{C-III}.\\
\emph{Proof of monotonicity.}
It is enough to show that $-\text{tr}\{\tilde{W}_\rho \rho\}$ is monotone under  free operations $\Lambda:\rho\rightarrow \sum_ip_iU_i\rho U_i^\dagger$. Hence, we have
\begin{eqnarray}
\label{eq-monotonicity_witness_based}
   - \text{tr} \{ \tilde{W}^{\Lambda({\rho})} \Lambda({\rho})    \}  = &&
    -\sum_i p_i \; \left[ \text{tr} \{ \tilde{W}^{\Lambda({\rho})} \rho_i    \} \right ]  \nonumber \\ \leq &&
    - \sum_i p_i\;  \underset{\tilde{W}}{\text{min}} ~~\text{tr} \{ \tilde{W} \rho_i    \} \nonumber \\ = &&
     -\sum_i p_i \; \text{tr}\{\tilde{W}^{\rho_i} \rho_i\} = 
    -\text{tr}\{\tilde{W}^{\rho} \rho\},\nonumber
    \end{eqnarray}
where we have used $U_i\rho U_i^\dagger=\rho_i$ and the last line follows from the fact that the witness-based NAS measure is invariant even under the global unitary operation.  \\
\textbf{Theorem 6.}
For pure states in $d\otimes d$, the witness-based measure of NAS, 
 $~~\mathcal{N}^W(\rho)=\frac{1}{d}.$
\begin{proof}
    Consider a pure state $\rho=\ket{\alpha}\bra{\alpha}$ in $d\otimes d$ which may be separable or entangled. To find $\mathcal{N}^W(\rho)$, we need to find the optimal unitary $U_\rho$ and the optimal entanglement witness operator $W_\rho$ for the state $\rho$. The minimum in Eq. (\ref{eq-witness_measure}) can be achieved when $U_\rho~:~U_\rho \rho U_\rho^\dagger=\rho_{\text{MES}}$, where $\rho_{\text{MES}}$ stands for the maximally entangled state (MES) in $d \otimes d$. Since the MES in $d\otimes d$ is a pure negative partially transposed (NPPT) entangled state, its optimal witness operator is $W_\rho=\ket{\phi}\bra{\phi}^{T_B}$, where $\ket{\phi}$ is the eigenvector corresponding to the smallest eigenvalue of $\rho_{\text{MES}}^{T_B}$\cite{TERHAL_2002}. Now, using the identity $\text{tr}\{\ket{\phi}\bra{\phi}^{T_B} \rho_{\text{MES}}\} = \text{tr}\{\ket{\phi}\bra{\phi} \rho_{\text{MES}}^{T_B}\}$, we can see that this is exactly the smallest eigenvalue of $\rho_{\text{MES}}^{T_B}$. Since $\rho_{\text{MES}}=\frac{1}{d}\sum_{i,j=1}^d\ket {ii}\bra{jj}$, we can find $$-\text{tr}\{W_\rho U_\rho \rho U_\rho^\dagger\} = 1/d,$$
    which leads to the same value for all pure states in $d\otimes d$.
\end{proof}
\textbf{Proposition 2.} 
The witness-based NAS measure of the modified Werner state ($\rho_W$) is given by $~~\mathcal{N}^W(\rho_W)=\text{max}\left[0,\frac{3p-1}{4}\right].$
\begin{proof}
    In $2\otimes2$, the optimal unitary operator which can generate entanglement can be characterized by only three parameters \cite{Vatan04},  given by 
    $$U=\exp\left[i(a_1 \sigma_x\otimes \sigma_x + a_2 \sigma_y\otimes \sigma_y+a_3 \sigma_z\otimes \sigma_z)\right].$$ Based on that unitary operator, we can find $\mathcal{N}^W(\rho)$. Also, in $2\otimes 2$ and $2\otimes 3$, every entangled state is NPPT  due to Peres-Horodecki criteria \cite{Peres96, horodecki96a}, and hence the optimal witness operator $W_{\rho_W}$ can be found by the prescription described in \textit{Theorem 6}. Therefore, we obtain
    \begin{eqnarray*}
        \mathcal{N}^W(\rho_W)&& =\text{max}\left\{ 0,~\underset{\{a_1,a_2,a_3\}}{\text{min}}\text{tr}\{U^\dagger W_{\rho_W} U \rho_W \}\right\}\nonumber\\&& = \text{max}\left[0,\frac{3p-1}{4}\right].\nonumber
    \end{eqnarray*}
\end{proof}
In Fig. \ref{fig:werner_quant}, we depict the witness-based NAS measure of the modified Werner state, $\mathcal{N}^W(\rho_W)$, with respect to $p$. We can see that like distance-based NAS measures, it also depends on the noise parameter, \(p\) and is also independent of the parameters involved in pure states of \(\rho_W\). As proven in \textit{Theorem 6}, 
at \(p=1\), it reaches to \(1/2\).

\section{Discussion}
\label{sec:conclu}

Developing a framework for a resource theory plays a pivotal role in quantum information science since it identifies states that are the essential ingredient for certain tasks to perform and recognizes operations that can be performed without any cost. For example, it has been demonstrated that entangled states are essential to gain a quantum advantage in a variety of protocols including quantum communication and measurement-based quantum computation. Hence, in the theory of entanglement, local operations and classical communication are considered free operations, by which entangled states cannot be created while separable states are free states. On the other hand, it has also been found that a class of separable states, known as absolutely separable states, cannot become entangled even with the help of global operations. 

In summary, we provided a resource theory to characterize the set of non-absolutely separable states (NAS) in which absolutely separable states and a mixture of global operations are free states and free operations. We proposed that NAS measures must satisfy certain requirements such as positivity, invariance with local unitary operations, monotonicity under mixture of global unitary operations, and convexity in order to be considered  ``good" measures. We employ two approaches for quantification -- (i) distance-based measures grounded on different metrics such as relative entropy, Bures, and Hilbert-Schmidt, and (ii) witness-based measures adopting witness operators used to detect non-absolutely separable states. We demonstrated that each measure  adheres to each requirement for a legitimate NAS measure. Since two completely different approaches are employed to quantify the resource content of NAS, no such obvious connections exist between these two kinds of measures. However, we noticed that the witness-based measures are a linear function in state parameter, whereas distance-based measures are not. Moreover, in the case of distance-based measures, we observed that the relative entropy-based measure shows higher value compared to the Bures distance-based measure.
\\
It is important to notice that the definitions of the NAS measures are independent of the number of parties comprising a state under evaluation. Specifically, our definition of distance-based measures of NAS does not consider the number of parties incorporated in the state. Hence, some of the main results (Theorems 1, 2 and 5) are valid even in the multipartite domain. On the other hand, the definition of the witness-based NAS measure is also applicable to multiparty cases because it does not presume the number of parties involved in a state. However, it should be noted that the criteria for being an absolutely separable state in a multipartite scenario are still not known and have not been thoroughly investigated due to their extreme mathematical complexity. Moreover, in the multipartite scenario, the notion of absolute separability is not well characterized as in the bipartite case, similar to the theory of entanglement. Therefore, in this work, we limited our analysis to the bipartite domain only.\\
We proved that all pure states in this picture possess an equal and maximum amount of resources. For given distance measures,  we obtained a compact form of NAS measures for all pure states which turn out to be an upper bound of an arbitrary density matrix. 
By taking modified Werner states, which are a mixture of a nonmaximally entangled state and white noise, we illustrated that NAS measures increase with the decrease of noise, regardless of the measures used. However, witness-based measures exhibit linear growth,  whereas distance-based measures do not. The resource theory presented in this paper, like other resource theories, may offer the opportunity to manipulate resources in novel ways, potentially leading to the creation of quantum devices.

\section*{Acknowledgment}
\label{sec:Ack}
We acknowledge the support from the Interdisciplinary Cyber-Physical Systems (ICPS) program of the Department of Science and Technology (DST), India, Grant No.: DST/ICPS/QuST/Theme- 1/2019/23. 
 
\section*{Appendix A: Bounds on the maximum eigenvalue}

 Consider a tuple of $2d$ numbers $\{\lambda_i^\downarrow\}_{i=1,2,...,2d}$ arranged in a non-increasing manner, i.e., $\lambda_i\geq\lambda_{i+1}$ with the condition $\sum_{i=1}^{2d} \lambda_i^\downarrow  =1$ and $\lambda_i\geq0$. We are looking for the bounds on $\lambda_1^\downarrow$ so that the condition  $\lambda_1^\downarrow - \lambda_{2d-1}^\downarrow - 2 \sqrt{\lambda_{2d-2}^\downarrow\lambda_{2d}^\downarrow }=0$ may hold. \\
 Note that the numbers $\{\lambda_i^\downarrow\}$ can be parameterized as
\begin{eqnarray*}
     \lambda_{1}^\downarrow  && = a_1, \nonumber \\ \lambda_2^\downarrow  &&
     = a_2 (1-a_1), \nonumber \\ \lambda_3^\downarrow   &&
     = a_3 (1-a_2)(1-a_1), \nonumber \\  &&
    \dots  \nonumber \\ \lambda_{2d-1}^\downarrow   &&
    = a_{2d-1} (1-a_{2d-2}) (1-a_{2d-3}) \dots (1-a_2)(1-a_1),  \nonumber \\ \lambda_{2d}^\downarrow   &&
     = (1-a_{2d-1})(1-a_{2d-2}) \dots (1-a_2)(1-a_1),
\end{eqnarray*}
where $\frac{1}{2d-i+1}\leq a_i \leq \text{min}\left[1,\frac{a_{i-1}}{1-a_{i-1}}\right]$ with $\frac{1}{2d}\leq a_1 \leq 1$.
To satisfy the equation, $\lambda_1^\downarrow - \lambda_{2d-1}^\downarrow - 2 \sqrt{\lambda_{2d-2}^\downarrow\lambda_{2d}^\downarrow }=0$, the necessary requirement is found to be 
\begin{eqnarray*}
    \frac{a_1}{1-a_1}\Big|_{\text{max}}&&= \prod_{i=2}^{2d-3}(1-a_{i,\text{min}})\\
   &&=\prod_{m=1}^{2d-4}\left(1-\frac{1}{2d-m}\right)\\
   &&=\frac{3}{2d-1}.
\end{eqnarray*}
Note that $\frac{a_1}{1-a_1}\big|_{\text{max}} = \frac{a_{1,\text{max}}}{1-a_{1,\text{max}}}$. Hence, we obtain $a_{1,\text{max}}=\frac{3}{2(d+1)}$. Again, $\lambda_1^\downarrow=a_1$, and hence the upper bound on $\lambda_1^\downarrow$ is $\frac{3}{2(d+1)}$. We now want to determine for whether the lower bound, i.e., $\frac{1}{2d}$, is achievable. Consider $\lambda_1^\downarrow=\frac{1}{2d}$, which immediately leads to all $\lambda_{i=2,3...2d}^\downarrow=\frac{1}{2d}$, for which the equality condition cannot be satisfied. Similarly, we can check the same for the upper bound by considering $\lambda_1^\downarrow=\frac{3}{2d+2}$ and $\lambda_{i=2,3...2d}^\downarrow=\frac{1}{2d+2}$, for which we can see that the condition is satisfied. Hence, $\frac{1}{2d}<\lambda_1^\downarrow\leq\frac{3}{2(d+1)}$.

\bibliographystyle{apsrev4-1}
\bibliography{AS}


\end{document}